\documentclass[12pt]{article}
\usepackage{epsf}
\usepackage{amsfonts} 
\usepackage{amsbsy} 

\newdimen\hsgraph \newdimen\vsgraph
\hsgraph=\hsize \advance\hsgraph by-1.2truein
\vsgraph=\vsize \advance\vsgraph by-1.5truein

\newcommand{\C}[1]{{\mathcal #1}}
\newcommand{\BS}[1]{{\boldsymbol #1}}
\newcommand{\R}[1]{{\mathrm #1}}

\newcommand{\beq}{\begin{equation}}
\newcommand{\eeq}{\end{equation}}
\newcommand{\bea}{\begin{eqnarray}}
\newcommand{\eea}{\end{eqnarray}}
\newcommand{\rf}[1]{(\ref{#1})} 

\newcommand{\half}{{1\over 2}}
\newcommand{\threehalves}{{3\over 2}}
\newcommand{\sevenhalves}{{7\over 2}}
\newcommand{\fivehalves}{{5\over 2}}
\newcommand{\threequarters}{{3\over 4}}
\newcommand{\third}{{1\over 3}}
\newcommand{\twothirds}{{2\over 3}}
\newcommand{\fourthirds}{{4\over 3}}
\newcommand{\fivethirds}{{5\over 3}}
\newcommand{\twofifths}{{2\over 5}}
\newcommand{\fourfifths}{{4\over 5}}
\newcommand{\fifth}{{1\over 5}}
\newcommand{\threefifths}{{3\over 5}}
\newcommand{\sevenfifths}{{7\over 5}}
\newcommand{\quarter}{{1\over 4}}

\newcommand{\nn}{\nonumber}
\newcommand{\Gbar}{{\overline G}}

\newcommand{\wbar}{{\overline w}}
\newcommand{\mymatrix}[1]{\underline{\underline{#1}}}
\newcommand{\mod}[1]{{\vert #1\vert}}
\newcommand{\expect}[1]{{\langle #1\rangle}}

\begin{document}
\topmargin 0pt
\oddsidemargin 5mm
\headheight 0pt
\topskip 0mm

\addtolength{\baselineskip}{0.20\baselineskip}

\pagestyle{empty}

\begin{flushright}
OUTP-99-44P\\
20th November 1999\\
revised 15th May 2000\\
hep-th/9911189
\end{flushright}

\begin{center}

\vspace{18pt}
{\Large \bf Peeling and  Multi-critical Matter
Coupled to Quantum Gravity}

\vspace{2 truecm}

{\sc Martin G. Harris\footnote{e-mail: martin.harris@mathengine.com.uk}
 \footnote{present address: MathEngine PLC, The Oxford Centre for Innovation, Mill Street,
Oxford OX2 0JX.} 
and John F. Wheater\footnote{e-mail: j.wheater1@physics.ox.ac.uk}}

\vspace{1 truecm}

{\em Department of Physics, University of Oxford \\
Theoretical Physics,\\
1 Keble Road,\\
 Oxford OX1 3NP, UK\\}

\vspace{3 truecm}

\end{center}

\noindent
{\bf Abstract.} We show how to determine the unknown functions arising when 
the peeling decomposition is applied to multi-critical matter coupled to 
two-dimensional quantum gravity and   compute  the loop-loop correlation functions. 
  The results that $\eta=2+2/(2K-3)$ and $\nu=1-3/2K$ 
agree with the slicing decomposition, and satisfy Fisher scaling.

\vfill
\begin{flushleft}
\end{flushleft}
\newpage
\setcounter{page}{1}
\pagestyle{plain}

\section{Introduction}

One of the outstanding problems in the theory of two-dimensional quantum 
gravity is the effect of matter fields on the Hausdorff dimension.
In models of discretized two-dimensional quantum gravity
we define the grand canonical partition function for an 
ensemble of graphs (which for the moment we assume are
triangulations) $\C G$ by
\beq \C Z(\mu)=\sum_{\R G \in \C G} e^{-\mu\mod {\R G}}w_{\R G}\label{1.1}\eeq
where $\mod {\R G}$ denotes the number of triangles in G, and $w_{\R G}$ the
partition function of any matter fields in the theory on  the graph G 
 (for an introduction to this material see for example \cite{book}).
To define the Hausdorff dimension \cite{kawai,AmbA} we first define the geodesic distance
 $d_{\R G}(i,j)$ between two links $i$ and $j$ as the minimum number of triangles
 which must be traversed to get from the centre of one link to the centre of the other. Then we introduce the two-point function
\beq \C H(r,\mu)=\sum_{\R G \in \C G} e^{-\mu\mod {\R G}}w_{\R G}\sum_{i,j\in \R G}
\delta(d_{\R G}(i,j)-r).\label{1.2}\eeq
We expect that $\C H$ has the asymptotic behaviour \cite{AmbA,AmbE}
\bea \C H(r,\mu)\sim& e^{-m(\mu) r},&\quad m(\mu) r>>1, \nn\\
\sim& r^{1-\eta_g},&\quad  m(\mu)^{-1} >> r>>1,\label{1.3}\eea
where,  as $\mu\to\mu_c$, the mass gap vanishes as
\beq m(\mu)\sim (\mu-\mu_c)^{\nu_g}.\label{1.4}\eeq
In general it is also convenient 
to consider a more general correlation function between boundary loops of
length $l_1$ and $l_2$; \rf{1.2} is essentially the correlator for minimum
length loops.
  Note  that it follows from
\rf{1.2} that 
\bea \sum_r  \C H(r,\mu)&=&\sum_{\R G \in \C G} e^{-\mu\mod {\R G}}\mod {\R G}^2
w_{\R G}\nn\\
&\sim& (\mu-\mu_c)^{-\gamma_{\rm str}}\label{1.5}\eea
where, in unitary theories,
 $\gamma_{\rm str}$ is the string susceptibility exponent,
and inserting the form \rf{1.3} we conclude that 
\beq \nu_g(2-\eta_g)=\gamma_{\rm str}\label{1.6}\eeq
which is the Fisher scaling relation.
At least in unitary theories the Hausdorff dimension
$d_H$ is given by $d_H\nu_g=1$ and has the geometrical meaning
 that in the continuum limit the average volume is related to the
geodesic size by $\expect{V}\sim R^{d_H}$ \cite{book}. 

Analytic calculations of the scaling behaviour of the correlation functions
\rf{1.2} were first done by means of the slicing decomposition introduced 
by Kawai et al \cite{kawai} and then 
somewhat later  Watabiki \cite{watabiki} introduced the peeling
decomposition. 
For pure gravity (ie $w_{\R G} =1$, $\gamma_{\rm str}=-1/2$) both peeling and slicing decompositions keep track of the geodesic distance and give the same results
which tell us directly that  the Hausdorff dimension of the ensemble is 4. 

 When matter fields are introduced the situation becomes more complicated.
 The time scale, usually called the string time $t$, introduced in the  decompositions that have been
formulated is no longer by construction the geodesic distance, nor indeed
are the time scales for different decompositions necessarily equivalent.
However the above discussion
of correlation functions can be repeated in terms of the string time $t$
instead of the geodesic distance $r$
 leading to another pair of exponents, $\eta$ and
$\nu$, which are also expected to satisfy the Fisher scaling relation.
 For example in the 
$c=-2$ model  the scaling with string time has been calculated completely by the peeling decomposition \cite{kristjansen}  with the result that $\nu=\half$  which would imply that $d_H=2$ if the string time and geodesic distance
 are proportional. In fact high precision numerical calculations \cite{anagnostopoulos} find $d_H=3.58\pm 0.04$ in agreement with  the formula \cite{watabikiA}
\beq d_H=2\frac{\sqrt{25-c}+\sqrt{49-c}}{\sqrt{25-c}+\sqrt{1-c}}\label{1.7}\eeq
derived using scaling arguments for diffusion in Liouville theory.
For unitary matter complete calculations have not been made but it seems 
that $\nu=\mod\gamma_{\rm str} /2$ \cite{ishibashi}; the implied value of $d_H$ is in contradiction with
the results of numerical simulations which suggest that $d_H$ is close to 4
\cite{anagnostopoulosA} but are not in particularly good agreement with
\rf{1.7} either.
It seems certain that when matter is present the string time and the geodesic
distance have different scaling dimensions but the relation between them
is unknown.

In this paper we will be concerned with the matrix models with a critical
point corresponding to  
the $(p,q)=(2,2K-1), K=2,3,4,\ldots$ multi-critical models coupled to
quantum gravity \cite{mcrit,mooreetal}.
For $K>2$ the ensemble of graphs $\C G$ now allows polygons with $\{4,\ldots 2K\}$ sides and there are $K$ independent coupling constants, with polygons
of order $\{6,10,14,\ldots\}$ having negative weights which makes the models
non-unitary. The partition function is again defined by
\beq \C Z(\mu)=\sum_{\R G \in \C G} e^{-\mu\mod {\R G}}w_{\R G}\label{1.A}\eeq
where now $\R G$ denotes the  number of polygons, $\mu$ 
is the coupling constant conjugate to the number of polygons,
 and $w_{\R G}$ depends on the
remaining $K-1$ coupling constants of the theory.  
$K=2$ corresponds to pure gravity; the second coupling constant which appears in this matrix model is conjugate to the
length of the boundary of the graph. 
Since there are now in general many couplings, there are many correlation functions
which are second derivatives with respect to the couplings so there are
many susceptibilites. We define  the susceptibility
\bea \chi_\mu\equiv\frac{\partial^2\C Z(\mu)}{\partial \mu^2}&=
&\sum_{\R G \in \C G} e^{-\mu\mod {\R G}}\mod {\R G}^2
w_{\R G}\nn\\
&\sim& (\mu-\mu_c)^{-\gamma}
\label{1.B}\eea
where the exponent $\gamma$ is known to take the value $-K^{-1}$ at the 
multicritical point.
The string time, $t_{\R G}(i,j)$, separating two links is now defined
 as the minimum number of polygons 
which must be traversed to get from the centre of one link to the centre of the other and the two-point function is defined as
\beq \C H(t,\mu)=\sum_{\R G \in \C G} e^{-\mu\mod {\R G}}w_{\R G}\sum_{i,j\in \R G}
\delta(t_{\R G}(i,j)-t).\label{1.C}\eeq
We expect (and shall confirm) that, if all the couplings except $\mu$ are set to their values at the multi-critical point,
 $\C H$ has the asymptotic behaviour 
\bea \C H(t,\mu)\sim& e^{-m(\mu) t},&\quad m(\mu) t>>1, \nn\\
\sim& t^{1-\eta},&\quad  m(\mu)^{-1} >> t>>1,\label{1.D}\eea
where,  as $\mu\to\mu_c$, the mass gap vanishes as
\beq m(\mu)\sim (\mu-\mu_c)^{\nu}.\label{1.E}\eeq
and that the Fisher scaling relation
(which may be obtained by similar manipulations as before)
\beq \nu(2-\eta)=\gamma\label{1.F}\eeq
is satisfied.
One could hold that $t$ is still the geodesic distance but
 this is slightly
problematic for large $K$; it implies that all sides of a given polygon, no 
matter how many sides it has,
are separated from one another by geodesic distance 1 and we shall argue
in section 6 that the continuum limit of $t$ is not a sensible continuum
geodesic distance.
These models have been analyzed using the slicing decomposition in the same way as pure
gravity
\cite{klebanovgubser} and they  have also been considered using the  peeling decomposition in the scaling limit but the discretized equations have not been
 solved completely \cite{watabiki,watabikiB}. 
In this paper we will examine  their peeling
decomposition in detail and explain how to solve completely the non-trivial
differential equations that arise.  We have two motivations for this;
to check whether the results are indeed the same as for slicing,
 and the intrinsic interest 
of the method of solution.

This paper is structured as follows. In section 2 we briefly describe the standard peeling calculation for pure gravity and then derive the evolution equation
for the multi-critical models. In section 3 we consider the $K=2$ case and show that it always gives the standard pure gravity results. Then in section 4 we show in detail how to calculate the $\eta$ exponent for $K=4$ and describe how the calculation extends to all higher even $K$. In section 5 we explain how to calculate $\nu$ for all even $K$ and in section 6 we give our conclusions.

\section{The Peeling Decomposition and Evolution Equations}

We start by reviewing the calculation in \cite{watabiki} for  the simplest pure gravity model which has matrix model potential
\beq U(\phi)=\half\phi^2-\third g\phi^3.\label{2.1}\eeq
The matrix model generates the dual graphs to the triangulations $\C G$
(see equation (1)) with $g=e^{-\mu}$ and $w_G=1$.
The Schwinger Dyson equation for connected Green's functions is 
obtained by marking one external line and pulling it out to expose the vertex to which it is attached \cite{Staudacher}, see fig.\rf{SD1}.
\begin{figure}[t]
\begin{center} \parbox{0.7\textwidth}{{\epsfxsize=0.7\textwidth \epsfbox{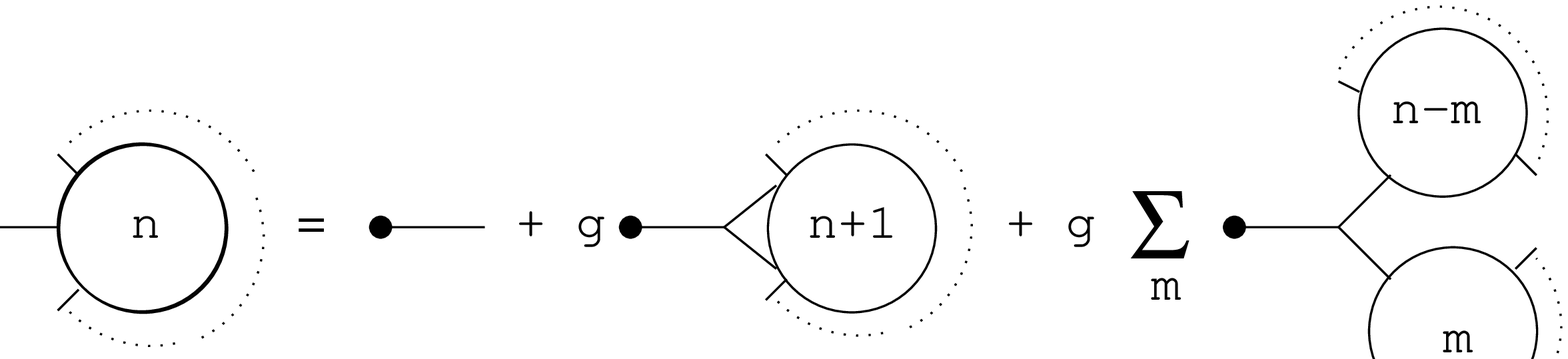}}}\end{center}
\caption{The Schwinger-Dyson equation for the potential \rf{2.1}.
 }\label{SD1}
\end{figure}
 In the peeling decomposition we assign a time variable to this process; a single iteration advances $t$ by an amount $1/n$ so that we obtain
\beq A_n(t+1/n)=\delta_{n,2}+g A_{n+1}(t)+g\sum_{m=1}^n A_m(t)A_{n-m+1}(t).\label{2.3}\eeq
We are interested in the loop-loop correlation function; suppose for the moment that at $t=0$ the entry loop is a one-loop and form the quantity
\beq G_n(t)=\frac{\delta A_n(t)}{\delta A_1(0)}\label{2.4}\eeq
which is the amplitude for an exit n-loop at time $t$ given an entry 1-loop at
time $t=0$. Differentiating \rf{2.3} we obtain 
\beq G_n(t+1/n)=g G_{n+1}(t)+ 2g\sum_{m=1}^n A_m(t)G_{n-m+1}(t).
\label{2.5}\eeq
If we restrict to spherical topology then $A_m(t)$ may be replaced by the disk amplitude for $m$ legs, $A_m$, because the branch can never rejoin the main tube
(see fig.\rf{tube}). 
\begin{figure}[t]
\begin{center} \parbox{0.6\textwidth}{{\epsfxsize=0.6\textwidth \epsfbox{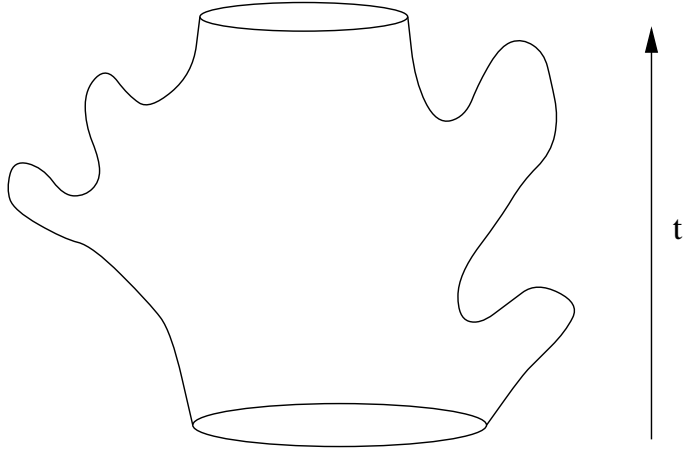}}}\end{center}
\caption{The two-loop function in spherical topology.}\label{tube}
\end{figure}
 The next step is to approximate the time by a continuous variable to obtain the evolution equation 
\beq G_n(t)+\frac{1}{n}\frac{dG_n(t)}{dt}=g G_{n+1}(t)+ 2g\sum_{m=1}^n A_m G_{n-m+1}(t)
\label{2.6}\eeq
with the initial condition that 
\beq G_n(0)=\frac{\delta A_n(0)}{\delta A_1(0)}=\delta_{n,1}.\label{2.7}\eeq
Note that by differentiating \rf{2.6} we can show iteratively that all the 
derivatives of $G_n(t)$ at $t=0$ are finite.
Defining the generating function
\beq G(t,x)=\sum_{n=1}^\infty x^n G_n(t)\label{2.8}\eeq
equation \rf{2.6} becomes
\beq -\frac{\partial G}{\partial t}=x\frac{\partial }{\partial x}
\left( \frac{ F(x)G}{x} \right)\label{2.9}\eeq
where
\beq F(x)=2gA(x)+g-x,\label{2.9a}\eeq
and the form of the disk amplitude
\beq A(x)=\sum_{n=1}^\infty A_n x^n\label{2.10}\eeq
is known \cite{ishibashi}. Note that the function $F(x)$ contains just the
universal scaling part of the amplitude $A(x)$; it is given by
\beq F(x)=(f-x)(1-8fg-4g x)^\half\label{2.12a}\eeq
where $f$ is the root of the cubic
\beq (1-8fg)f^2-g^2=0\label{2.12b}\eeq
which is positive and vanishes as $g\to 0$. It is then straightforward to
solve the evolution equation which gives
\beq G(t,x)=\frac{x}{F(x)}U(t+J(x))\label{2.14}\eeq
where
\beq \frac{dJ}{dx}=\frac{1}{F(x)}\label{2.15}\eeq
and
the function $U(y)$ is fixed by the initial conditions to satisfy
\beq 1=\frac{1}{F(x)}U(J(x)).\label{2.15a}\eeq
This leads to the scaling behaviour at large $y$
\beq U(y)\simeq(\mu-\mu_c)^\threequarters\frac{\cosh(\mu-\mu_c)^\quarter  y}
{\sinh^3(\mu-\mu_c)^\quarter  y}\label{2.16}\eeq
where we have suppressed various constant factors;  so we deduce $\nu=\quarter$ and $\eta=4$ in agreement with the Fisher scaling relation \cite{AmbA}.

Now we turn  to the multi-critical models for which we will use the notation of
reference \cite{HarrisWeiss}.  The  potential for
the $K$-th multi-critical model is 
given by 
\beq U(\Phi)=\Phi^2+\sum_{p=1}^Kg^K_p\Phi^{2p}\label{2.17}\eeq
where the couplings at the multicritical point are
\beq g^K_p=(-1)^{p-1}\frac{K!(p-1)!}{(K-p)! 2p!}-\half \delta_{p,1}.\label{2.18}\eeq

\begin{figure}[t]
\begin{center} \parbox{0.7\textwidth}{{\epsfxsize=0.7\textwidth \epsfbox{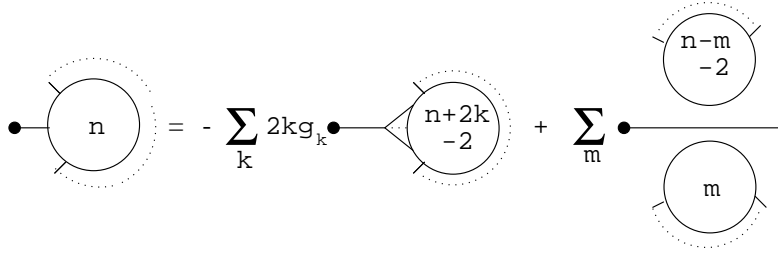}}}\end{center}
\caption{The Schwinger-Dyson equation for disconnected graphs of
the multicritical potential \rf{2.17}.
 }\label{SD2}
\end{figure}

This time we will deal with \emph{disconnected} graphs.
The evolution equation  follows from the Schwinger-Dyson equation,
shown in fig.\ref{SD2},  just as in the pure gravity case;
 we obtain the evolution equation
\beq A_n(t+1/n)=\sum_{j=0}^{n-2} A_{n-j-2}(t)A_j(t)-2\sum_{k=1}^Kkg^K_kA_{n+2k-2}(t).\label{2.20}
\eeq
Proceeding as before, but assuming that the entry loop is an $m$-loop and defining
\beq G_{n,m}(t)=\frac{\delta A_n(t)}{\delta A_m(0)},\label{2.21}\eeq
we find
\beq  G_{n,m}(t)+\frac{1}{n}\frac{d  G_{n,m}(t)}{dt}=
2\sum_{j=0}^{n-2} G_{j,m}(t) A_{n-j-2}-\sum_{k=1}^K 2kg^K_k  G_{n+2k-2,m}(t)\label{2.22}\eeq
with the initial condition that 
\beq G_{n,m}(0)=\delta_{n,m}.\label{2.22a}\eeq
Again, note that by differentiating \rf{2.22} we can show iteratively that all the 
derivatives of $G_{n,m}(t)$ at $t=0$ are finite.

Defining the generating function
\beq G_m(x,t)=\sum_{n=1}^\infty x^n G_{n,m}(t)\label{2.23}\eeq
we obtain the partial differential equation
\bea \frac{\partial G_m}{\partial t}&=& x\frac{\partial }{\partial x}
\left\{G_m\left(2x^2A(x)-1-\sum_{k=1}^K\frac{ 2kg^K_k}{x^{2k-2}}\right)\right\}
\nn\\&&\qquad-\sum_{k=1}^K 2kg^K_k\sum_{j=1}^{2k-3}G_{j,m}(t)(2k-2-j)x^{j-2k+2}.\label{2.24}\eea
At this stage we should make several remarks. Firstly that one reason for
dealing with the disconnected graphs is technical convenience; the required
disk amplitudes are known and take a simple form \cite{HarrisWeiss}, and the 
structure of the evolution equations is similar to the $\phi^3$ case.
However it is also very clear that because the graph ensemble is disconnected
there is no direct correspondence between the string time $t$ and a geodesic distance -- the latter can only be sensibly defined on connected graphs.
These equations are more difficult to solve than the pure gravity example
reviewed  above because of the presence of the  \emph{a priori} unknown 
  functions $G_{j,m}(t)$ which have to be determined by the self-consistency and analyticity  properties of the solutions.
We will work our way through the problem in a number of steps.
 First we will study the solution in the $K=2$ case and show that it gives the standard pure gravity results.

\section{Universality in the $K=2$ case}

It is simpler to work at the multi-critical points initially and to compute the $\eta$ exponent directly at the critical point; we leave the $\nu$ exponent 
for section 5. For the $K=2$ multi-critical model we have
\beq\frac{\partial G_m}{\partial t}=x\frac{\partial }{\partial x}
\left( G_m(2x^2A(x)-1-2g^2_1-4g^2_2x^{-2}\right) - 4g^2_2 x^{-1}G_{1,m}(t)\label{3.1}\eeq
where $g^2_1=1/2$, $g^2_2=-1/12$ and  $A(x)$ is given by
\beq A(x)=\frac{1}{x^2}+\frac{1}{6x^4}\left((1-4x^2)^{\threehalves}-1\right).\label{3.2}\eeq
Thus $G_m$ satisfies
\beq\frac{\partial x^{-1} G_m}{\partial t}=\frac{\partial }{\partial x}
\left(\frac{ G_m(1-4x^2)^{\threehalves}}{3x^2}\right) +\frac{1}{3x^2}G_{1,m}(t)\label{3.3}\eeq
with the initial condition
\beq G_m(0,x)=x^m.\label{3.3a}\eeq

The potential \rf{2.17} is an even function of the fields and therefore
$G_{n,m}(t)$, which is the amplitude for an entrance $m$-loop and exit $n$-loop,
can only be non-zero of $n$ and $m$ are both odd or both even. Thus if $m$ is odd there is an unknown function on the r.h.s. of \rf{3.3} whereas if $m$ is even there 
is no such problem; we will consider the even and odd cases separately.

\subsection{Even $m$}
When $m$ is even $G_{1,m}(t)=0$ and we can solve \rf{3.3} immediately to obtain
\beq G_m(t,x)=\frac{x}{F(x)}U(t+J(x))\label{3.1.1}\eeq
where
\bea F(x)=\frac{ (1-4x^2)^{\threehalves}}{3x},\nn\\
\frac{dJ}{dx}=\frac{1}{F(x)},\nn\\
J(x)=\threequarters\left((1-4x^2)^{-\half}-1\right),\label{3.1.2}\eea
and the function $U(y)$ is determined by the initial condition \rf{3.3a}
\beq x^m=\frac{x}{F(x)}U(J(x)).\label{3.1.3}\eeq
Thus we obtain 
\beq U(y)\sim\frac{1}{y^3}\label{3.1.4}\eeq
for large $y$ and hence  
\beq G_{n,m}(t)\sim\frac{1}{t^3}\label{3.1.5}\eeq
for even $n$ and $m$ so that $\eta=4$ as expected for all these
amplitudes.

\subsection{Odd $m$}
When $m$ is odd $G_{1,m}(t)\ne 0$ and the  presence of the unknown function 
on the r.h.s. of \rf{3.3} complicates matters; however this is more typical
of the general multi-critical models than the even $m$ case and so we shall
study it in some detail. Although the differential equation can of course still
be solved in the $t$ domain it is more convenient to work with the Laplace
transformed correlation functions
\bea \Gbar_{n,m}(s)=\int_0^\infty G_{n,m}(t) e^{-st} dt\nn\\
\Gbar_m(s,x)=\frac{1}{x}\int_0^\infty G_m(t,x) e^{-st} dt \label{3.4}.\eea
Taking the Laplace transform of \rf{3.3} we obtain the equation
\beq s\Gbar_m(s,x)-x^{m-1}=\frac{\partial }{\partial x}\left(\Gbar_m(s,x) F(x)\right) +
\frac{1}{3x^2}\Gbar_{1,m}(s).\label{3.6}\eeq
Integrating this differential equation we find
\beq \Gbar_m(s,x)=\frac{1}{F(x)}\left(
\frac{\Gbar_{1,m}(s)}{3x}-e^{sJ(x)}\int_0^x dy e^{-sJ(y)}\left(y^{m-1}-
\frac{s \Gbar_{1,m}(s)}{(1-4y^2)^{\threehalves}}\right)\right).\label{3.8}\eeq
\vspace{1 truecm}

The function $\Gbar_m(s,x)$ has a power series expansion in $x$
\beq \Gbar_m(s,x)=\sum_nx^{n-1} \Gbar_{n,m}(s)\label{3.11}\eeq
which we expect from \rf{3.8} and \rf{3.1.2} to have finite radius of convergence $\half$; within the radius of convergence the coefficients are the Laplace transforms of the correlation functions. Unless the $G_{n,m}(t)$ grow faster
than exponentially at large $t$, something we do not expect, their Laplace transforms $\Gbar_{n,m}(s)$  will have an asymptotic series representation at large $s$. This series can be obtained by successive integration by parts of the 
definition \rf{3.4}; as we observed in section 2, $G_{n,m}(t)$ and all its (finite order) derivatives are finite at $t=0$  so we obtain the formal series
\beq \Gbar_{n,m}(s) =\frac{1}{s}\sum_{k=0}\frac{\alpha_k}{s^k}, \quad \alpha_0=\delta_{n,m}.\label{3.12}\eeq
(Of course the $\alpha_k$ depend on $n$ and $m$ but we will always suppress
such dependence for clarity.)
We will now show that imposing the \emph{condition} that the coefficients in the $x$ expansion of  $\Gbar_m(s,x)$ behave like \rf{3.12} at large $s$ is
sufficient to fix the unknown function $\Gbar_{1,m}(s)$.

It is more convenient to impose the condition on the integral of $\Gbar_m(s,x)$,
\beq\int_0^x \Gbar_m(s,y) dy= \frac{x^m}{sm}-\frac{1}{s}e^{sJ(x)}\int_0^x dy
e^{-sJ(y)}\left(y^{m-1}-
\frac{s \Gbar_{1,m}(s)}{(1-4y^2)^{\threehalves}}\right)\label{3.13}.\eeq
Using \rf{3.11}, \rf{3.12} and \rf{3.13} the consistency condition is that 
\beq g_m=e^{sJ(x)}\int_0^x dy
e^{-sJ(y)}\left(y^{m-1}-
\frac{s \Gbar_{1,m}(s)}{(1-4y^2)^{\threehalves}}\right)\sim\frac{1}{s}.\label{3.14}\eeq
That is to say the small $x$, large $s$, expansion of $g_m$ contains no
terms $ O(1)$ or higher in $s$.  Substituting \rf{3.12} and expanding the 
exponential in its Taylor series (which is of course allowed for all values of the argument)  we get
\beq g_m=\int_0^x dy\sum_{p=0}^\infty\frac{s^p}{p!}\left(J(x)-J(y)\right)^p
\left(y^{m-1}-
\frac{1}{(1-4y^2)^{\threehalves}}\sum_{k=0}\frac{\alpha_k}{s^k}\right).\label{3.15}\eeq
For $x<\half$ \rf{3.15} clearly has a power series expansion in $x$ so the requirement that there are no $O(1)$ terms in $s$ becomes
\beq C_0=\int_0^x dy\left(y^{m-1}-
\frac{1}{(1-4y^2)^{\threehalves}}\sum_{p=0}^\infty\left(J(x)-J(y)\right)^p\frac{\alpha_p}{p!}\right)=0.\label{3.16}\eeq
First we show that this constraint alone 
implies that all higher powers of $s$ vanish as well; for $O(s)$ we get
\beq C_1=\int_0^x dy\left(J(x)-J(y)\right)\left(y^{m-1}-
\frac{1}{(1-4y^2)^{\threehalves}}\sum_{p=0}^\infty\left(J(x)-J(y)\right)^p\frac{\alpha_p}{p+1!}\right)=0,\label{3.17}\eeq
and observe that
\beq\frac{dC_1}{dx}=\frac{dJ}{dx} C_0=0\label{3.18}
\eeq
provided $x<\half$. It follows that $C_1(x)$ is a constant; but  $C_1(0)=0$
therefore
\beq C_1(x)=0\label{3.19}.\eeq
Any pair $C_{k}(x)$ and  $C_{k+1}(x)$, with $k>0$ are related in the same way
and so  it is straightforward to proceed inductively to show that all $C_{k>0}(x)$ are
zero.

Taking the Laplace transform of \rf{3.16} with respect to $J$ we find that 
\beq \int_0^\infty\frac{x(J)^m}{m}e^{-sJ}\,dJ= \Gbar_{1,m}(s)\int_0^\infty
\frac {e^{-sJ}}{3x(J)}\, dJ\label{3.20}\eeq
where $x(J)$ is obtained by inverting \rf{3.1.2}. Thus $ \Gbar_{1,m}(s)$ is determined; making the change of variables $x=\half\tanh\phi$ and integrating by parts we get 
\beq \Gbar_{1,m}(s)=\frac{ \int_0^\infty\tanh^{m-1}\phi\; e^{-\threequarters s\cosh\phi\,}\R{{sech}}^2\phi\, d\phi}{2^{m-1}s\int_0^\infty\cosh\phi\; e^{-\threequarters s\cosh\phi}\, d\phi}. \label{3.21}\eeq
It is clear that $ \Gbar_{1,m}(s)$ is positive for all real, positive $s$ and 
straightforward to check that its second derivative with respect to $s$ 
diverges logarithmically as $s\to 0$; it follows that 
\beq G_{1,m}(t)\sim\frac{1}{t^3}\label{3.22}\eeq
at large $t$. The remaining integrals in \rf{3.8} yield functions which are
analytic in $s$ in the neighbourhood of the origin provided $x<x_c$. Thus we 
can conclude that 
\beq G_{n,m}(t)\sim\frac{1}{t^3}\label{3.22a}\eeq
for all odd $n$ and $m$. So the result $\eta=4$ is universal for all amplitudes $G_{n,m}(t)$ in agreement with every other calculation for pure gravity.

In fact $ \Gbar_{1,m}(s)$ can be fixed more simply by examining \rf{3.8}. As
$x\uparrow x_c$, $J(x)$ diverges and hence $\Gbar_m(s,x)$ grows faster than
exponentially in $s$, which is impossible, unless
\beq \int_0^\half dx e^{-sJ(x)}\left(x^{m-1}-\frac{s\Gbar_{1,m}(s)}
{(1-4x^2)^{\threehalves}}\right)=0 \label{3.23}\eeq
which is the same condition as \rf{3.21}. We have explored the more indirect route
because this will help in the multi-critical case.

\section{The Multi-critical Models}

The multi-critical evolution equation \rf{2.24} may be written
\bea \frac{\partial x^{-1} G_m(x,t)}{\partial t}&=& \frac{\partial }{\partial x}
\left\{x^{-1}G_m(x,t) F(x)\right\}\nn\\
&&\qquad-\sum_{k=1}^K 2kg^K_k\sum_{j=1}^{2k-3}G_{j,m}(t)(2k-2-j)x^{j-2k+1}\label{4.1}\eea
where, taking $A(x)$ from \cite{HarrisWeiss}, we have
\bea F(x)&\equiv& x\left(2x^2A(x)-1-\sum_{k=1}^K\frac{ 2kg_k}{x^{2k-2}}\right)\nn\\
&=& -2Kg^K_K x^{3-2K}(1-4x^2)^{K-\half}.\label{4.2}\eea
The solution to \rf{4.1} takes the form \cite{CourantHilbert}
\bea x^{-1}G_m&=&\frac{1}{F(x)}U_0(t+J(x))\nn\\
&&\; -\sum_{k=1}^K 2kg^K_k\sum_{j=1}^{2k-3}\frac{(2k-2-j)}{F(x)}
\int_0^t d\tau G_{j,m}(\tau)U_{jk}(t-\tau+J(x))\label{4.3}\eea
where, as before,
\beq J(x)=\int_0^x \frac{dy}{F(y)}=\sum_{r=0}^{K-2}
h^K_r (1-4x^2)^{r-K+\threehalves}-h^K_{K-1}\label{4.4}\eeq
and the coefficients $h^K_r$ are easy to compute. The function $U_0(y)$ is fixed by the initial conditions
\beq x^{m-1}=\frac{1}{F(x)}U_0(J(x))\label{4.5}\eeq
and the functions $U_{jk}(y)$ by requiring that \rf{4.3} is a solution of
\rf{4.1},
\beq  x^{j-2k+1}=\frac{1}{F(x)}U_{jk}(J(x)).\label{4.6}\eeq
Because of the unknown functions $G_{j,m}(t) $, \rf{4.3}  is not of course a complete solution. Note that $J(x)$ is always an even function which diverges as $\vert x\vert\to \half$ and which is positive for even $K$ and negative for odd $K$. To determine the critical behaviour from the properties at large $t$ we need to know the behaviour of the functions
$U_0(\tau)$ and  $U_{jk}(\tau)$ for large positive argument. For even $K$ we see
  by considering \rf{4.5} and \rf{4.6} as $\vert x\vert\to \half$ that
\bea U_0(\tau)&\stackrel{\tau\to\infty}{\sim}&\tau^{-\frac{2K-1}{2K-3}},\nn\\
 U_{jk}(\tau)&\stackrel{\tau\to\infty}{\sim}&\tau^{-\frac{2K-1}{2K-3}}.\label{4.6a}\eea
However for odd $K$ only the large negative argument  behaviour is determined. This phenomenon always occurs in  calculations for the multi-critical models.
 Extrapolating the solution to positive time leads to a singularity at finite time; it seems to us quite likely that this is an artefact of the truncation of the original finite
difference equation \rf{2.20} into a first order differential equation \rf{2.22} and that the
solution may be stabilised by higher derivative terms.  From now on we will concentrate on the even $K$ models and start by studying the $K=4$ case in detail.

\subsection{$K=4$, even $m$}
As for the $K=2$ model we will consider the cases of even and odd $m$ separately. For even $m$ we have
\beq  \frac{\partial x^{-1} G_m(x,t)}{\partial t}= \frac{\partial }{\partial x}
\left\{x^{-1}G_m(x,t) F(x)\right\}+G_{2,m}(t)\left(\frac{4}{35x^5}
-\frac{4}{5x^3}\right)+\frac{2}{35x^3}G_{4,m}(t)\label{4.7}\eeq
with
\beq F(x)=\frac{(1-4x^2)^{\sevenhalves}}{35x^5}.\label{4.8}\eeq
Taking the Laplace transform gives
 \beq s\Gbar_m(s,x)-x^{m-1}=\frac{\partial }{\partial x}\left(\Gbar_m(s,x) F(x)\right) +
\Gbar_{2,m}(s)\left(\frac{4}{35x^5}
-\frac{4}{5x^3}\right)+ \frac{2}{35x^3}\Gbar_{4,m}(s)\label{4.9}\eeq
where the transformed correlation functions are defined as in \rf{3.4}; integrating \rf{4.9} we
obtain
\bea	\Gbar_m(s,x)&=&\frac{1}{F(x)}\Bigg\{\frac{1}{35x^4}\Gbar_{2,m}(s)
+ \frac{1}{35x^2}\Gbar^C_{4,m}(s)	+ \frac{1}{35}\Gbar^C_{6,m}(s)	\nn\\
&&\quad+e^{sJ(x)}\int_0^xdy e^{-sJ(y)}
 \bigg(-y^{m-1}\nn\\
&&\qquad +\frac{s}{35y^5F(y)}\Big(y^5\Gbar^C_{6,m}(s)+y^3\Gbar^C_{4,m}(s)+y\Gbar_{2,m}(s)\Big)\bigg)\Bigg\}\label{4.10}\eea
where we have introduced the combinations
\bea \Gbar^C_{4,m}(s)&=&\Gbar_{4,m}(s)-14\Gbar_{2,m}(s),\nn\\
\Gbar^C_{6,m}(s)&=&70\Gbar_{2,m}(s)-14\Gbar_{4,m}(s)+\Gbar_{6,m}(s).\label{4.10a}\eea
Note the appearance of $\Gbar_{6,m}(s)$ in \rf{4.10}; this happens because $F(x)$ is singular at $x=0$ which makes the evaluation of the limits of integration slightly non-trivial.
As before it is convenient to deal with 
\beq\int_0^x\Gbar_m(s,y)dy=\frac{x^m}{sm}+\frac{g_m}{s}\label{4.12}\eeq
where now
\bea g_m(s,x)&=&e^{sJ(x)}\int_0^xdy e^{-sJ(y)}
 \Bigg\{-y^{m-1}+\nn\\
&&\quad\frac{s}{35y^5F(y)}\left(y^5\Gbar^C_{6,m}(s)+y^3\Gbar^C_{4,m}(s)+y\Gbar_{2,m}(s)\right)\Bigg\}\label{4.13}\eea
and we deduce that
\beq g_m\sim\frac{1}{s}\label{4.14}\eeq
in order to fulfil the constraint that $\Gbar_m(s,x)$ has the correct small $x$, large $s$, expansion. 
 In fact we can immediately determine $\Gbar_{6,m}(s)$ as explained at the end of section 2 by requiring that $\Gbar_m(s,x)$ does not grow faster than any exponential of $s$ as $x\uparrow\half$ which
implies that
\beq g_m(s,\half)=0\label{4.11}\eeq
and so $\Gbar_{6,m}(s)$ is related to
$\Gbar_{2,m}(s)$ and $\Gbar_{4,m}(s)$ which still have to be determined.

Defining
\bea \Gbar^C_{6,m}(s) &=&\frac{1}{s}\sum_{k=0}\frac{\alpha_k}{s^k},\nn\\
\Gbar^C_{4,m}(s) &=&\frac{1}{s}\sum_{k=0}\frac{\beta_k}{s^k},\nn\\
\Gbar_{2,m}(s)&=&\frac{1}{s}\sum_{k=0}\frac{\gamma_k}{s^k},\label{4.15}\eea
(of course the $\alpha$, $\beta$ and $\gamma$ coefficients depend on $m$
but we have suppressed this to avoid clutter)  and substituting into \rf{4.13} we obtain the $O(s^0) $ constraint
\beq C_0=-
\int_0^x dy\left\{y^{m-1}-
\frac{1}{35y^5F(y)}\sum_{k=0}^\infty\frac{(J(x)-J(y))^k}{k!}
(y^5\alpha_k+y^3\beta_k+y\gamma_k)\right\}=0.\label{4.16}\eeq
Identical manipulations to the $K=2$ case show that $C_0=0$ ensures $C_{k>0}=0$ also. There are in fact three separate constraints hidden in \rf{4.16} which are sufficient to determine the $\alpha_k$, $\beta_k$ and $\gamma
_k$ coefficients. To see this we proceed by writing everything in terms of $J$.  $J(x)$ is an even function with a small $x$ expansion
\beq J(x)=\frac{35}{6}x^6+\frac{245}{4}x^8+\ldots\label{4.16a}\eeq
so it follows that, by reverting the series,
\beq\frac{x(J)^m}{m}=(J^\third)^{m/2}\sum_{k=0}^\infty a_k(J^\third)^k\label{4.17}\eeq
where we take the real positive cube root of $J$
(remember that $m$ is even and again we suppress the $m$ dependence on $a_k$). Similarly we have 
\bea\int_0^x y dy \frac{1}{35y^5F(y)}(J(x)-J(y))^k
&=&\int_0^J\frac{ dJ'}{35 x(J')^4}(J-J')^k\nn\\
&=&\int_0^J dJ'(J-J')^k J'^{-\twothirds}
	\sum_{l=0}^\infty b_l(J'^\third)^l\nn\\
&=&\sum_{l=0}^\infty b_l J^{k+(l+1)/3}\frac{\Gamma(k+1)\Gamma(\frac{l+1}{3})}
{\Gamma(k+\frac{l+1}{3}+1)},\label{4.18}\eea
and
\beq\int_0^x y^3 dy \frac{1}{35y^5F(y)}(J(x)-J(y))^k
=\sum_{l=0}^\infty c_l J^{k+(l+2)/3}\frac{\Gamma(k+1)\Gamma(\frac{l+2}{3})}
{\Gamma(k+\frac{l+2}{3}+1)}.\label{4.19}\eeq
We will not need explicit expressions for the coefficients $a_l$, $b_l$ and
$c_l$.  Substituting \rf{4.17}, \rf{4.18} and \rf{4.19} into $C_0$ we obtain
\bea C_0(J)=&&-
(J^\third)^{m/2}\sum_{k=0}^\infty a_k(J^\third)^k+\frac{1}{35}\sum_{k=0}^\infty\frac{\alpha_kJ^{k+1}}{(k+1)!}\nn\\
&&+\sum_{k=0}^\infty\sum_{l=0}^\infty\gamma_k b_l J^{k+(l+1)/3}\frac{\Gamma(\frac{l+1}{3})}{\Gamma(k+\frac{l+1}{3}+1)}\nn\\
&&+\sum_{k=0}^\infty\sum_{l=0}^\infty\beta_k c_l J^{k+(l+2)/3}\frac{\Gamma(\frac{l+2}{3})}{\Gamma(k+\frac{l+2}{3}+1)}.\label{4.20}\eea
Letting $N$ be an integer, \rf{4.20}  yields three conditions corresponding to terms $O(J^N)$, $O(J^{N+\third})$ and $O(J^{N+\twothirds})$ respectively.
The corresponding coefficients are easily extracted from \rf{4.20} and we find
\beq \frac{\alpha_{N-1}}{35}-a_{3N-m/2}\Gamma(N+1)+\sum_{L=1}^N\gamma_{N-L}b_{3L-1}\Gamma(L)+\sum_{L=1}^N\beta_{N-L}c_{3L-2}\Gamma(L)=0,\label{4.21}\eeq
\beq-a_{3N+1-m/2}\Gamma(N+\fourthirds)+\sum_{L=0}^N\gamma_{N-L}b_{3L}\Gamma(L+\third)+\sum_{L=1}^N\beta_{N-L}c_{3L-1}\Gamma(L+\third)=0,\label{4.22}\eeq
\beq -a_{3N+2-m/2}\Gamma(N+\fivethirds)+\sum_{L=0}^N\gamma_{N-L}b_{3L+1}\Gamma(L+\twothirds)+\sum_{L=0}^N\beta_{N-L}c_{3L}\Gamma(L+\twothirds)=0.\label{4.23}
\eeq
Now suppose that $\gamma_0\ldots\gamma_{N-1}$ and $\beta_0\ldots\beta_{N-1}$ are known; from \rf{4.21} we can obtain $\alpha_0\ldots\alpha_{N-1}$.
Then \rf{4.22} contains only these known coefficients together with $\gamma_N$ which is thus determined. Now \rf{4.23} determines $\beta_N$. 
Knowing $\gamma_0\ldots\gamma_{N-1}$ and $\beta_0\ldots\beta_{N-1}$ we now determine $\alpha_N$ from \rf{4.21}. Proceeding in this way all the coefficients, and thus the right hand sides of \rf{4.15}, can be obtained iteratively.
To analyse the asymptotic behaviour of the correlation functions we need to know the behaviour of $\Gbar_{2,m}(s)$ etc at small $s$. Resumming the relations
\rf{4.21}, \rf{4.22}, and \rf{4.23}, and using the integral representation
\beq \frac{\Gamma(n)}{s^n}=\int_0^\infty d\xi \xi^{n-1}e^{-\xi s},\label{4.24}\eeq
we get
\bea 0= \frac{\Gbar^C_{6,m}(s)}{35s}&-&\int_0^\infty d\xi e^{-\xi s}\sum_{N=1}^\infty a_{3N-m/2}\xi^N
+\Gbar_{2,m}(s)\int_0^\infty d\xi e^{-\xi s}\sum_{N=1}^\infty b_{3N-1}\xi^{N-1}\nn\\
&+&\Gbar^C_{4,m}(s)\int_0^\infty d\xi e^{-\xi s}\sum_{N=1}^\infty c_{3N-2}\xi^{N-1},\label{4.25}
\eea
\bea 0&=&-\int_0^\infty d\xi e^{-\xi s}\sum_{N=0}^\infty a_{3N+1-m/2}\xi^{N+\third}
+\Gbar_{2,m}(s)\int_0^\infty d\xi e^{-\xi s}\sum_{N=0}^\infty b_{3N}\xi^{N-\twothirds}\nn\\
&+&\Gbar^C_{4,m}(s)\int_0^\infty d\xi e^{-\xi s}\sum_{N=1}^\infty c_{3N-1}\xi^{N-\twothirds},\label{4.26}\eea
\bea 0&=&-\int_0^\infty d\xi e^{-\xi s}\sum_{N=0}^\infty a_{3N+2-m/2}\xi^{N+\twothirds}
+\Gbar_{2,m}(s)\int_0^\infty d\xi e^{-\xi s}\sum_{N=0}^\infty b_{3N+1}\xi^{N-\third}\nn\\
&+&\Gbar^C_{4,m}(s)\int_0^\infty d\xi e^{-\xi s}\sum_{N=0}^\infty c_{3N}\xi^{N-\third}.\label{4.27}\eea
These three equations determine the unknown functions that we need.

Now consider the equation
\beq J(x)=\xi\label{4.28}\eeq
where $\xi$ is real and positive; it can be rewritten as the quintic equation
for $u\equiv x^2$
\beq (1+\frac{24}{7}\xi)^2(1-4u)^5-(30u^2-10u+1)^2=0.\label{4.32}\eeq
When $\xi=0$ it reduces to 
\beq -4u^3(10-95u+256u^2)=0\label{4.33}\eeq
which has a complex conjugate pair of roots
\beq u=\frac{95\pm i 9\sqrt{15}}{512},\label{4.34}\eeq
and three roots vanishing when $\xi=0$; any integer power of these roots, which we denote
by $u_{1,2,3}$, has a series expansion
\bea (u_1(\xi))^M&=&(\xi^\third)^M\sum_{n=0}^\infty D_{Mn}(\xi^\third)^n,\nn\\
{}(u_2(\xi))^M&=&(\omega\xi^\third)^M\sum_{n=0}^\infty D_{Mn}(\omega\xi^\third)^n,\nn\\
{}(u_3(\xi))^M&=&(\omega^2\xi^\third)^M\sum_{n=0}^\infty D_{Mn}(\omega^2\xi^\third)^n, \label{4.29}
\eea
where $\omega$ is a complex cube root of unity and $\xi^\third$ is the real
cube root of $\xi$.

Now take linear combinations of  \rf{4.25}- \rf{4.27} with the coefficients being different powers of $\omega$ to obtain
\beq 0=\int_0^\infty d\xi e^{-s\xi}\left\{u_i(\xi)^{\half m}
-\frac{1}{35}\left(\frac{\Gbar_{2,m}(s)}{u_i(\xi)^2}+\frac{\Gbar^C_{4,m}(s)}{u_i(\xi)}
+\Gbar^C_{6,m}(s)\right)\right\}\label{4.30}\eeq
with $i=1,2,3$.

\begin{figure}[ht]
\begin{center}\parbox{0.4\textwidth}{{\epsfxsize=0.4\textwidth \epsfbox{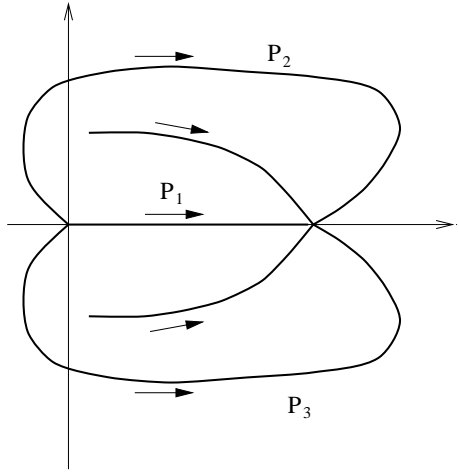}}}\end{center}
\caption{The flow of the roots of \rf{4.32} in the complex $u$ plane; the arrows denote the direction of increasing $\xi$.}\label{rootsK4}
\end{figure}

As $\xi\to\infty$ all the roots of \rf{4.32} converge on $u=\quarter$ like
\beq u=\quarter\left(1-\sigma\frac{c}{\xi^\twofifths}+\ldots\right)\label{4.35}\eeq
where $\sigma $ is a fifth root of unity and $c$ is a constant.
 The flow of the roots in the complex plane as $\xi$ varies is shown in fig.\ref{rootsK4}. We see that $u_1$ flows into the $\xi=\infty $ point as \rf{4.35} with $\sigma=1$; and that the complex conjugate pair
$u_{2,3}$ flow into the pair with $\sigma=e^{\mp i\frac{4\pi}{5}}$. 
Note that there are no degenerate roots of \rf{4.28} in the interval $\xi=(0,\infty)$ (this follows from the fact that $F$ is regular in this interval).

 Now define
the function $\C J$ by
\beq \C J(u=x^2)=J(x)\label{4.36a}\eeq	
 and make the change of variable
\beq \xi=\C J(u_i)\label{4.37}\eeq
in \rf{4.30}.
The constraint equations then take their final form
\beq 0=\int_{ P_i} du \frac{d\C J}{du}e^{-s\C J(u)}\left\{u^{\half m}
-\frac{1}{35}\left(\frac{\Gbar_{2,m}(s)}{u^2}+\frac{\Gbar^C_{4,m}(s)}{u}
+\Gbar^C_{6,m}(s)\right)\right\}\label{4.38}\eeq
where the contours $ P_i$ in the complex $u$-plane are shown in fig.\rf{rootsK4}.
Note that the $i=1$ constraint is the same as \rf{4.11} after an integration by
parts and that these equations are guaranteed to have a unique solution by the argument immediately following \rf{4.21}-\rf{4.23}.

\begin{figure}[ht]
\begin{center}\parbox{0.4\textwidth}{{\epsfxsize=0.4\textwidth \epsfbox{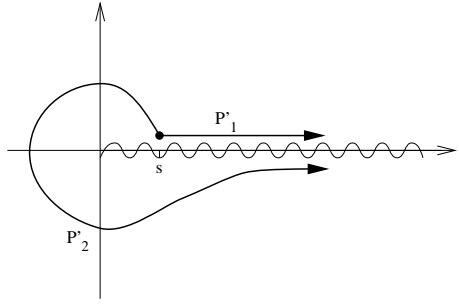}}}\end{center}
\caption{Integration contours $ P_{1,2}'$ for  \rf{4.42} in the complex $z$ plane. All contours start at $z=s$ and go out to infinity; the cut is for the 
fractional powers of $z$.  The contour  $ P_{3}'$ is simply the complex conjugate of $ P_{2}'$.}\label{zpathK4}
\end{figure}

\begin{figure}[ht]
\begin{center}a)~~\parbox{0.4\textwidth}{{\epsfxsize=0.4\textwidth \epsfbox{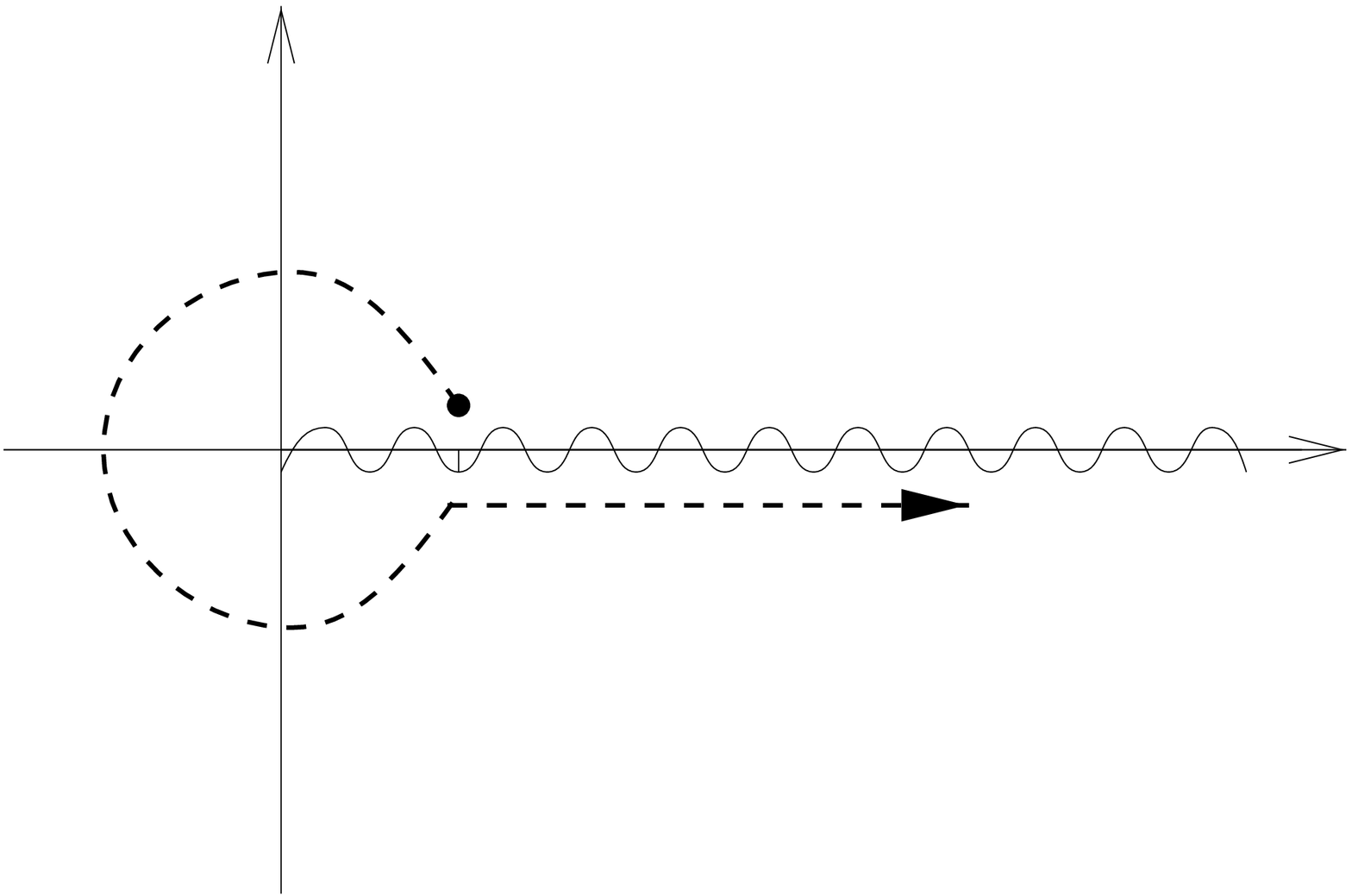}}}\hfil b)~~\parbox{0.4\textwidth}{{\epsfxsize=0.4\textwidth \epsfbox{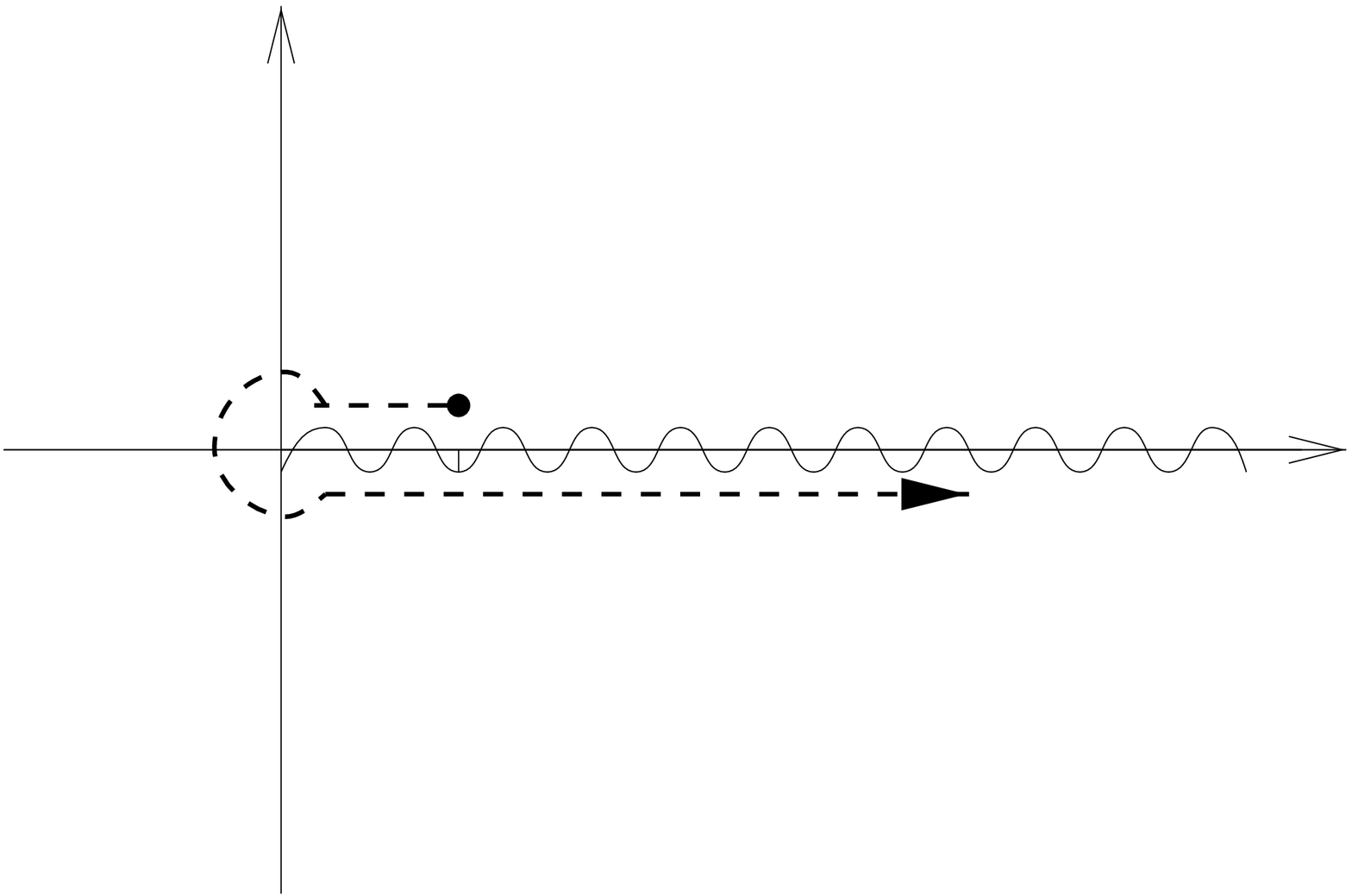}}}\end{center}
\caption{a) Distortion of the integration contour $ P_{2}'$. If the integrand has
an integrable singularity at the origin then the circular part of the new contour can also be collapsed onto the branch cut as shown in b).
 }\label{dzpathK4}
\end{figure}

To find the asymptotic large $t$ dependence of the correlation functions
it suffices to find the leading non-trivial $s$-dependence of their Laplace
tranforms at small $s$; for this purpose we need the integrals
in \rf{4.38}
up to and including $O(s^\fifth)$. Replacing $\C J$ by its explicit form
we need
\beq I^{\C F}_i=\frac{35}{2}\int_{ P_i} du\,u^2(1-4u)^{-7/2}\C F(u)
\exp(-s(A(1-4u)^{-\fivehalves}+B(1-4u)^{-\threehalves}+C(1-4u)^{-\half}))\label{4.40}\eeq
where $\C F(u)=\left\{u^{-2},u^{-1},u^{\half m}\right\}$, $A=7/64$, $B=-35/96$ and $C=35/64$ and we have dropped a factor
$\exp(7s/24)$ which cancels in \rf{4.38}.
Now make the change of variables
\beq z=s(1-4u)^{-\fivehalves}.\label{4.41}\eeq
Then all the integrals in \rf{4.37} are linear combinations of
\beq \C H_k=s^{(2k-7)/5}\int_{ P_i'}dz z^{-2(k-1)/5}\exp(-Az)
\exp\left(-(Bs^\twofifths z^\threefifths +Cs^\fourfifths z^\fifth)\right)
\label{4.42}\eeq
where $k=1,2,\ldots$. The contours
$ P_i'$ in the complex $z$-plane are shown in fig.\ref{zpathK4}. 
The second exponential in \rf{4.42} is then expanded in its Taylor series,
and the $ P_{2,3}'$ contours collapsed onto the branch cut as shown in fig.\ref{dzpathK4}; 
all the integals then become elementary. The equations \rf{4.38} can then be solved to give
\bea \Gbar_{2,m}(s)&=&\frac{35}{2^{7+m}}(m+2)+\frac{175}{6.2^{7+m}}
\frac{\Gamma(\fourfifths)}{\Gamma(\fifth)}(m^2+6m+8)(1+2\cos\frac{4\pi}{5})
\left(\frac{7s}{64}\right)^\twofifths+O(s^\fourfifths),\nn\\
 \Gbar_{4,m}(s)&=&\frac{35}{2^{6+m}}(3m-2)+\frac{175}{2^{7+m}}
\frac{\Gamma(\fourfifths)}{\Gamma(\fifth)}(m^2+6m+8)(1+2\cos\frac{4\pi}{5})
\left(\frac{7s}{64}\right)^\twofifths+O(s^\fourfifths),\nn\\
 \Gbar_{6,m}(s)&=&\frac{35}{2^{6+m}m}(15m^2-50m+64)\nn\\&&+\frac{875}{2^{7+m}}
\frac{\Gamma(\fourfifths)}{\Gamma(\fifth)}(m^2+6m+8)(1+2\cos\frac{4\pi}{5})
\left(\frac{7s}{64}\right)^\twofifths+O(s^\fourfifths).\label{4.43}
\eea
These expressions have the expected properties; they are positive and decreasing functions of $s$ for small positive $s$. It follows from  \rf{4.43} that
$G_{2,m}(t)$,  $G_{4,m}(t)$ and   $G_{6,m}(t)$ all have the same
 asymptotic 
behaviour at large $t$ namely
\beq  G_{j,m}(t)\sim t^{-\sevenfifths}\label{4.44}\eeq
for $j=2,4,6$
and hence the exponent $\eta=12/5$. To determine the behaviour of the higher correlation functions it is simplest to return to \rf{4.10}. 
All the integrals yield functions of $s$ which are analytic in some neighbourhood of the origin provided $x<1/2$ and so we can conclude that every coefficient of the $x$ expansion of $G_m(x,t)$ behaves the same way and that \rf{4.44} is valid for 
all (even) $j$ and $m$.

\subsection{$K=4$, odd $m$}

The method is very similar to the even $m$ case. After Laplace transforming the 
evolution equation and solving for $\Gbar_m(s,x)$ we obtain
\bea\lefteqn{	 \Gbar_m(s,x)=\frac{1}{F(x)}\Bigg\{\frac{1}{35x^5}\Gbar_{1,m}(s)+
\frac{1}{35x^3}\Gbar^C_{3,m}(s)+\frac{1}{35x}\Gbar^C_{5,m}(s)	}\nn\\&+&
e^{sJ(x)}\int_0^x dy e^{-sJ(y)}\Big(-y^{m-1}\nn\\&+&\frac{s}{35y^5F(y)}
\bigg(y^4\Gbar^C_{5,m}(s)+y^2\Gbar^C_{3,m}(s)+\Gbar_{1,m}(s)\bigg)\Big)\Bigg\}
\label{4.45}\eea
The consistency conditions can be cast in the form
\beq 0=\int_{ P_i} du \frac{d\C J}{du}e^{-s\C J(u)}\left\{u^{\half m}
-\frac{1}{35}\left(\frac{\Gbar_{1,m}(s)}{u^\fivehalves}+\frac{\Gbar^C_{3,m}(s)}{u^\threehalves}
+\frac{\Gbar^C_{5,m}(s)}{u^\half}\right)\right\}\label{4.45a}\eeq
This time we need the integrals
\beq I^{\C F}_i=\frac{35}{2}\int_{ P_i} du\frac{u^2}{u^\half}(1-4u)^{-7/2}
\{u^{-2},u^{-1},1,u^{\half(m+1)}\}
e^{-s\C J(u)}\label{4.46}\eeq
To the required order in $s$ these can be calculated by expanding the
$u^{-\half}$ factor  about $u=\quarter$; then term by term the integrals are just the $\C H_k$. The resulting expressions for the singular terms in $s$ are simple but the constants appear as infinite sums; fortunately the constants are only needed for the  sub-leading $s$ dependence so this method suffices (the reason for this is explained in the next sub-section). The final expressions for the correlation functions are very similar to those for even $m$
\bea \Gbar_{1,m}(s)&=&\frac{35}{2^{8+m}m}(m^2+4m+3)\nn\\&+&\frac{175}{6.2^{8+m}m}
\frac{\Gamma(\fourfifths)}{\Gamma(\fifth)}(m^3+9m^2+23m+15)(1+2\cos\frac{4\pi}{5})
\left(\frac{7s}{64}\right)^\twofifths+O(s^\fourfifths),\nn\\
 \Gbar_{3,m}(s)&=&\frac{35}{2^{7+m}m}(3m^2+4m+1)\nn\\&+&\frac{175}{2^{8+m}m}
\frac{\Gamma(\fourfifths)}{\Gamma(\fifth)}(    m^3+9m^2+23m+15  )(1+2\cos\frac{4\pi}{5})
\left(\frac{7s}{64}\right)^\twofifths+O(s^\fourfifths),\nn\\
 \Gbar_{5,m}(s)&=&\frac{35}{2^{7+m}m}(15m^2-20m+29)\nn\\&&+\frac{875}{2^{8+m}}
\frac{\Gamma(\fourfifths)}{\Gamma(\fifth)}( m^3+9m^2+23m+15 )(1+2\cos\frac{4\pi}{5})
\left(\frac{7s}{64}\right)^\twofifths+O(s^\fourfifths).\nn\\&&\label{4.47}
\eea
Thus we can conclude that the exponent $\eta$ always takes the value
$12/5$ in the $K=4$ model.

\subsection{General even $K$}
The method is very similar to the  $K=4$ case. After Laplace transforming the 
evolution equation and solving for $\Gbar_m(s,x)$ we obtain
\bea\lefteqn{	 \Gbar_m(s,x)=\frac{1}{F(x)}\Bigg\{
-2Kg^K_K\sum_{p=2}^{2K-2}x^{1-p}\Gbar^C_{2K-p-1,m}(s)
	}\nn\\&+&
e^{sJ(x)}\int_0^x dy e^{-sJ(y)}\Big(-y^{m-1}\nn\\&-&\frac{2Kg^K_Ks}{y^{2K-3}F(y)}
\sum_{p=1}^{2K-2}y^{p-1}\Gbar^C_{p,m}(s)
\Big)\Bigg\}
\label{4.60}\eea
where
\beq \Gbar^C_{2K-p-1,m}(s)=\sum_{k=2}^K\frac{kg^K_k}{Kg^K_K}\Gbar_{2k-p-1,m}(s).\label{4.61}\eeq
Note that for given $m$ (either odd or even) half of the $\Gbar^C_{p,m}(s)$ in
\rf{4.60} are automatically zero so there are $K-1$ undetermined functions.
Proceeding as before  we next examine the roots of the equation
\beq J(x)=\xi\label{4.62}\eeq
which can be rewritten as a degree $2K-3$ polynomial equation for $u\equiv x^2$
taking the generic form
\beq (1+a\xi)^2(1-4u)^{2K-3}-(1+b_1u+\ldots+b_{K-2}u^{K-2})^2=0\label{4.63}\eeq
where $a$, $b_{1},\ldots b_{K-2}$ are constants. There are $K-1$ roots which vanish at $\xi=0$ and $K-2$ roots which do not; the $K-1$ vanishing roots correspond to the $K-1$ functions which have to be determined. As $\xi\to\infty$ all the roots converge on $u=1/4$ like
\beq u=\quarter\left(1-\sigma\frac{c}{\xi^\frac{2}{2K-3}}+\ldots\right)\label{4.64}\eeq
where $c$ is a constant and $\sigma$ is a $(2K-3)$rd root of unity.
The roots which vanish at $\xi=0$ flow into 
\beq\sigma=\exp\left( i\frac{4\pi n}{2K-3}\right),\quad n=-\frac{K}{2}+1,\dots,
0,\ldots \frac{K}{2}-1.\label{4.64a}\eeq
  From now on we will use the integer $n$ to label the roots.
The consistency conditions become
\beq 0=\int_{ P_n} du \frac{d\C J}{du}e^{-s\C J(u)}\left\{u^{\half m}
+2Kg^K_K\sum_{p=1}^{2K-2}\frac{\Gbar^C_{p,m}(s)}{u^{(2K-2-p)/2}}\right\}\label{4.65}\eeq
where the $K-1$ paths $ P_n$ follow in the complex plane the roots
which vanish at the origin.  Note that in the case of $K=4$ the conditions
for even $m$, \rf{4.38}, and for odd $m$, \rf{4.45a}, can be merged into the form
\rf{4.65}; for any fixed $m$ we have $K-1$ equations for $K-1$ unknowns.

Restricting ourselves to even $m$ all the integrals in \rf{4.65} can be written as linear combinations  of
\beq\C K_p=\int_{ P_n} du (1-4u)^{p-K+\half}\exp\left(sh^K_{K-1}-s\sum_{r=0}^{K-2}
h^K_r (1-4u)^{r-K+\threehalves}\right).\label{4.66}\eeq
Note that the factor $\exp(sh^K_{K-1})$ cancels out in \rf{4.65} so from now on we drop it. After the substitution
\beq z=s(1-4u)^{-K+\threehalves}\label{4.67}\eeq
we obtain
\beq 2(2K-3)\C K_p=s^{-1+\frac{2p}{2K-3}}\int_{\C P'_n}dz
z^{-\frac{2p}{2K-3}}\exp\left(-h^K_0z-\sum_{r=1}^{K-2}
h^K_r s^{\frac{2r}{2K-3}}z^{1-\frac{2r}{2K-3}}\right)\label{4.68}\eeq
where the contours $ P'_n$ encircle the origin $n$ times before heading
off to real positive infinity, see fig.(7a).
\begin{figure}[ht]
\begin{center}a)~~\parbox{0.4\textwidth}{{\epsfxsize=0.4\textwidth
 \epsfbox{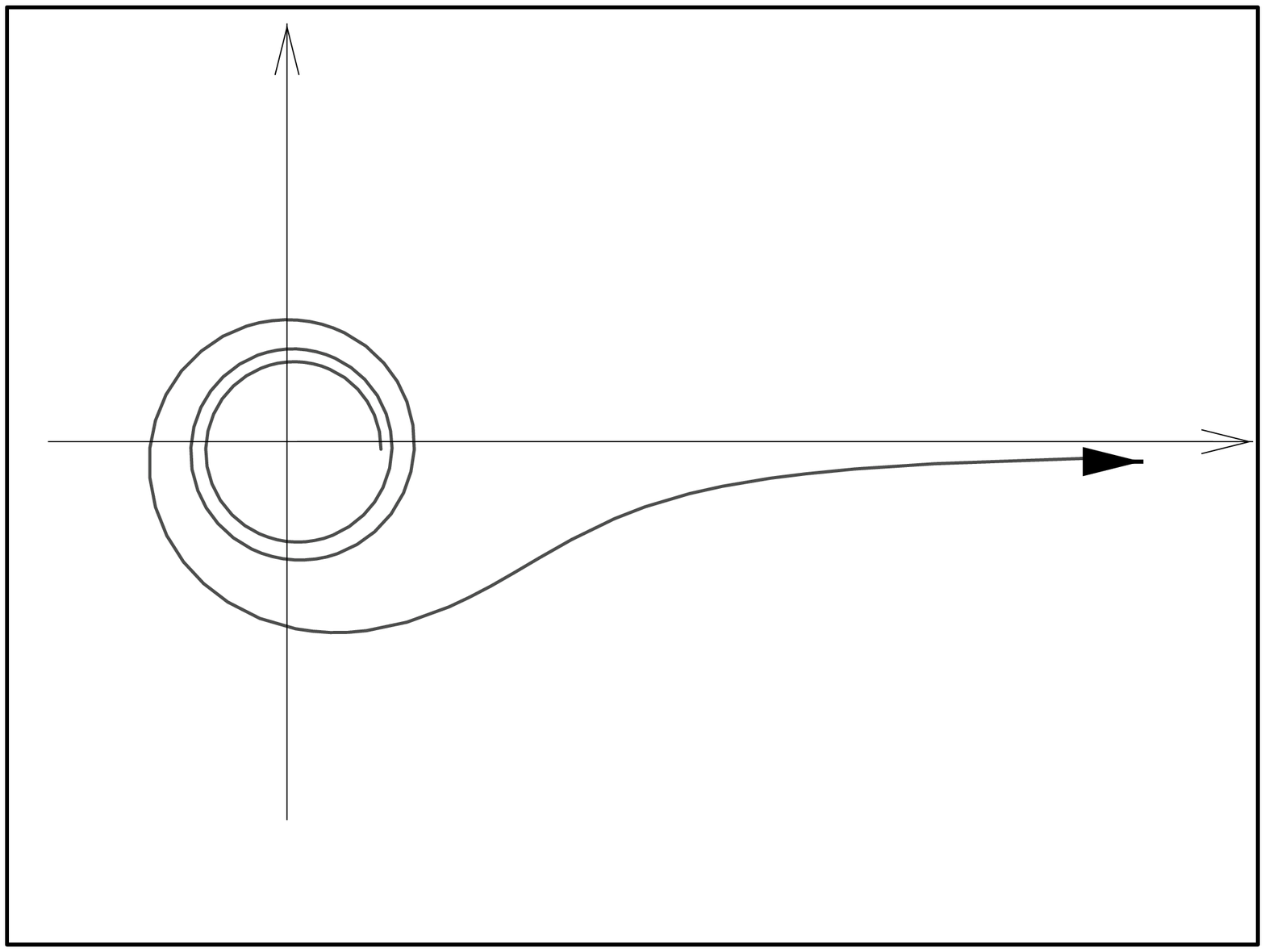}}}\hfil b)~~
\parbox{0.4\textwidth}{{\epsfxsize=0.4\textwidth
 \epsfbox{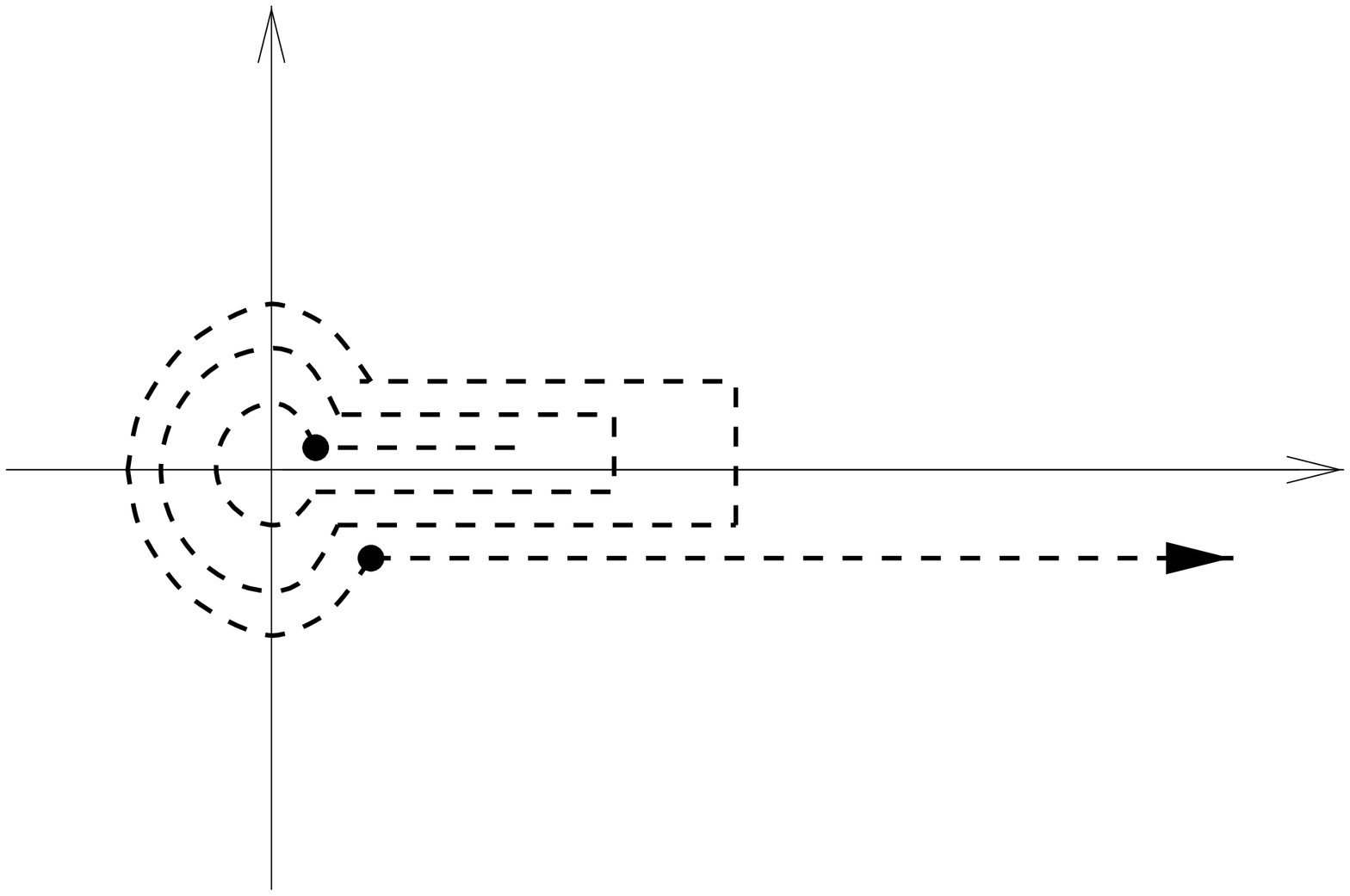}}}
\end{center}
\caption{a) A contour $ P'$ with $n=3$ and b) the deformed version of it. The branch cut on the positive real axis is suppressed for clarity.}\label{spiral}
\end{figure}
 As we will demonstrate shortly all these integrals are needed up to and including the first positive power
of $s$ and so  they can be separated into three classes;
\begin{enumerate}
\item $p<K-1$; the integrals are divergent as $s\to 0$ but the singularities of the integrand are integrable at $z=0$ so the contour can be deformed as shown
in fig.(7b).
 The integral from the starting point to the first blob is given by 
\beq 2(2K-3)\C K_p^{fin}=\frac{2K-3}{2p-2K+3} +O(s).\label{4.69}\eeq
The portion of the contour between the two blobs gives no contribution 
and the leg stretching out to infinity gives
\bea 2(2K-3)\C K_p^{div}&=&
(se^{-i2\pi n})^{-1+\frac{2p}{2K-3}}\nn\\&&\int_0^\infty dR R^{-\frac{2p}{2K-3}}
\exp\left(-h^K_0R-\sum_{r=1}^{K-2}
h^K_r R \left(\frac{se^{-i2\pi n}}{R}\right)^{\frac{2r}{2K-3}}\right).\nn\\
&&\label{4.70}\eea
To obtain the integral up to the desired order the second exponential factor
can be Taylor expanded and then integrated term by term.
\item $p=K-1$; after an integration by parts the above construction can be used
and we get
\beq 2(2K-3)\C K_{K-1}=(2K-3)\left(1-(h^K_0 se^{-i2\pi n})^{\frac{1}{2K-3}}
\Gamma\left(\frac{2K-4}{2K-3}\right)\right)+\ldots.\label{4.71}\eeq
\item $p>K-1$; the leading non-analytic term is of higher order than we
need to consider so
\beq 2(2K-3)\C K_{p}=\frac{2K-3}{2p-2K+3}.\label{4.72}\eeq
\end{enumerate}

Now observe that in all these integrals $s$ always appears with a factor
$e^{-i2\pi n}$ and that there are no other phase factors. Thus all of the
constraint equations \rf{4.65} can be obtained from the $n=0$ case by making the replacement $s\to se^{-i2\pi n}$. They can thus be written in the form
\beq \C F_0(se^{-i2\pi n})+\sum_{p=1}^{K-1}\C F_p(se^{-i2\pi n})\Gbar^C_{2p,m}(s)=0\label{4.73}\eeq
where
\beq\C F_p(w)=w^{-1}\left(\sum_{q=0}^{K-1}f_{p,q} w^{\frac{2q}{2K-3}}
+f_{p,K} w+\ldots\right).\label{4.74}\eeq
We have truncated the expansion of $\C F_p(w)$ in
 anticipation of the following.
The equations \rf{4.73} can now be written in the matrix form
\beq\mymatrix\Omega\, \mymatrix D {\BS f}
+s^{\frac{1}{2K-3}}f_{0,K-1}{\BS \omega}
+\mymatrix\Omega\, \mymatrix D\,\mymatrix f{\BS G}
+s^{\frac{1}{2K-3}}\mymatrix{\tilde{f}}{\BS G}=0\label{4.75}
\eeq
where
\bea \mymatrix{\Omega}_{pq}&=&\exp\left(i2\pi(-K/2+p)(-1+2q/(2K-3))\right),\nn\\
\mymatrix{D}_{pq}&=&\delta_{pq}s^{2(p-1)/(2K-3)-1},\nn\\
{\BS f}_p&=&f_{0,p-1}+f_{0,K}s\delta_{p,1},\nn\\
{\BS \omega}_p&=&\exp\left(i2\pi(-K/2+p)/(2K-3)\right),\nn\\
\mymatrix{f}_{pq}&=&f_{p,q-1}+f_{q,K}s\delta_{p,1},\nn\\
\mymatrix{\tilde{f}}_{pq}&=&\exp\left(i2\pi(-K/2+p)/(2K-3)\right)f_{q,K-2}\nn\\
{\BS G}_p&=&\Gbar^C_{2p,m}.\label{4.76}\eea
Now $\mymatrix{D}$ is clearly non-singular and it is straightforward to
check that $ \mymatrix{\Omega}$ is non-singular.
There does not seem to be any simple way of writing the elements of
$\mymatrix{f}$ for general $K$ but we expect it too  is non-singular.
Then  the leading order solution for $\BS G$
is a constant vector; furthermore the next 
term in the solution is $O(s^{2/(2K-3)})$. We have already shown by explicit solution that this is indeed what happens for $K=4$; using Maple we have also 
checked it for $K=6,8,10$.
It follows that 
\beq \eta=\frac{4(K-1)}{2K-3}\label{4.77}\eeq
for the correlation functions $\Gbar^C_{p,m}, p=2,4,\ldots 2(K-1)$. The relationship \rf{4.61} between $\Gbar^C_{p,m}$ and $\Gbar_{p,m}$ is non-singular so this
conclusion applies also to $\Gbar_{p,m}, p=2,4,\ldots 2(K-1)$. Finally we note as usual that the integrals in \rf{4.60} are analytic functions of $s$ in the 
neighbourhood of the origin provided $x<1/2$ so the conclusion extends to all the correlation functions.

\section{The $\nu$ exponent}

To find $\nu$  we need to study the scaling behaviour as the 
multi-critical point is approached and to do this consider the modified
couplings
\bea \tilde g^K_1&=&g^K_1\nn\\
\tilde g^K_p&=&(1-\Delta) g^K_p,\quad p\ge 2.\label{5.1}\eea
In the graphical expansion  the power of $1-\Delta$
is the number of vertices in the graph so $\Delta$ is related to $\mu$ in
(8) by
\beq 1-\Delta=\exp(\mu-\mu_c).\eeq 
 By definition the multi-critical point is attained as $\Delta\to 0$. The disk amplitude $A(x)$ can still be calculated by exploiting the connection with topological gravity \cite{HarrisWeiss}; the topological potential is
\beq V(z)=\Delta-Kz\Delta+z+(1-\Delta)(1-z)^K\label{5.2}\eeq
and the disk amplitude
\beq A(\Delta,x)=-\oint_C\frac{dz}{2\pi i}(1-4zx^2)^{-\half}(1-V'(z))\log\frac{z-V(z)}{z}\label{5.3}\eeq
where the contour encircles the branch cut of the logarithm. The branch points
are at $z=0$ and 
\beq z=\frac{z_c}{4}\simeq 1-\epsilon_K\Delta^{1/K}\label{5.4}\eeq
where $\epsilon_K$ is a constant.
 Collapsing the contour onto the cut gives
\beq  A(\Delta,x)=K\int_0^{z_c/4}dy (1-4yx^2)^{-\half}\left(\Delta+(1-\Delta)(1-y)^{K-1}\right)
\label{5.5}\eeq
Integrating we find that 
\bea F(x)&\equiv& x\left(2x^2A(\Delta,x)-1-\sum_{k=1}^K\frac{ 2k\tilde g^K_k}{x^{2k-2}}\right)\nn\\
&=& -xK\Delta(1-z_cx^2)^\half-x(1-\Delta)  \sum_{k=1}^K\frac{ 2k g^K_k}{x^{2k-2}}(\epsilon_K\Delta^{1/K})^{K-k}(1-z_cx^2)^{k-\half}
\nn\\&&\label{5.6}\eea
which reproduces \rf{4.2} when $\Delta=0$.

At small $x$, even for finite $\Delta$ we still get the leading behaviour
\beq F(x)\sim x^{3-2K}\label{5.7}\eeq
which implies that 
\beq J(x)=\int_0^x\frac{dy}{F(y)}\sim x^{2K-2}.\label{5.8}\eeq
Thus the considerations described in detail for $K=4$ in section 4 up to equation \rf{4.27} go through as before -- the only difference is that all the various coefficients are now functions of $\Delta$. However the contours $ P_i$ are
modified; when we make the change of variable \rf{4.37}
\beq \xi=\C J(u_i)\label{5.9}\eeq
the $\xi=\infty$ endpoint of the contour occurs at the value of $u_i$ where 
$\C J$ diverges. The points where $\C J$ diverges are of course determined by the zeros of $F$.  Defining $w$ through
\beq x^{-2}=z_c+\epsilon_K\Delta^{1/K}w\label{5.10}\eeq
we get using \rf{5.6}
\bea \lefteqn{\C J(w)\equiv\C J\left(u=(z_c+\epsilon_K\Delta^{1/K}w)^{-1}\right)=}\nn\\
&&\frac{(\epsilon_K\Delta^{1/K})^\half}{2}\int_\infty^w
\frac{w^{-\half}(z_c+\epsilon_K\Delta^{1/K}w)^{-\half}}
{K\Delta+2(1-\Delta)(\epsilon_K\Delta^{1/K})^{K-1}\sum_{k=1}^Kkg^K_kw^{k-1}}\,dw.
\label{5.11}\eea
The term $K\Delta$ in the denominator is sub-leading; its only effect is to shift the zeros by an amount $O(\Delta^{1/K})$  and we discard it to obtain
\beq \C J(w)=
\frac{(\epsilon_K\Delta^{1/K})^{\threehalves-K}}{4(1-\Delta)}\int_\infty^w
\frac{w^{-\half}(z_c+\epsilon_K\Delta^{1/K}w)^{-\half}}
{\sum_{k=1}^Kkg^K_kw^{k-1}}\,dw.
\label{5.12}\eeq
Note that the denominator has precisely $K-1$ simple zeroes; one is real and positive and the others come in complex conjugate pairs. Each zero is the end-point of one of the $K-1$ contours, see fig.\rf{splitrts}.
\begin{figure}[ht]
\begin{center}
\parbox{0.4\textwidth}{{\epsfxsize=0.4\textwidth
 \epsfbox{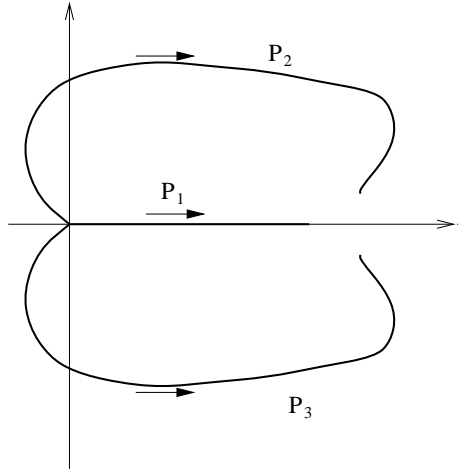}}}
\end{center}
\caption{The contours for $K=4$ at finite $\Delta$.}\label{splitrts}
\end{figure}
 Since we will only need the leading scaling behaviour we can also approximate $ (z_c+\epsilon_K\Delta^{1/K}w)$ by $z_c$ in \rf{5.12}  so that 
\bea \C J(w)&=&
\frac{(\epsilon_K\Delta^{1/K})^{\threehalves-K}}{4(1-\Delta)(-Kg^K_K)}
\int^\infty_w\frac{w^{-\half}}{ \prod_{k=1}^{K-1}(w-\wbar_k)}dw\nn\\
&=&\frac{(\epsilon_K\Delta^{1/K})^{\threehalves-K}}{4(1-\Delta)(-Kg^K_K)}
\sum_{k=1}^{K-1}R_k\int^\infty_w\frac{w^{-\half}}{w-\wbar_k}dw
\label{5.13}\eea
where $\wbar_k$ is a zero of the denominator of \rf{5.12} and $R_k$ the 
corresponding residue.

  Now suppose that $w$ lies on the contour $ P_l$
which terminates at  $\wbar_l$; when $w$ is close to the end-point we get
\beq \C J(w)=
\frac{(\epsilon_K\Delta^{1/K})^{\threehalves-K}}{4(1-\Delta)(-Kg^K_K)}
\left(\frac{R_l}{\sqrt{\wbar_l}}\log\frac{\sqrt{w}+\sqrt{\wbar_l}}
{\sqrt{w}-\sqrt{\wbar_l}}+Q_l\right)\label{5.14}\eeq
where $Q_l$ is the accumulated finite contribution from all the other poles in \rf{5.13}. Note that the branch cut for the square roots is taken down the negative real axis so that $Re\sqrt{\wbar_l}>0$ for all $l$.  Inverting \rf{5.14}
we find that 
\beq w=\wbar_l\left(\frac{T_l\exp\left(\Delta^{1-3/2K}S_l\C J\right)+1}
{T_l\exp\left(\Delta^{1-3/2K}S_l\C J\right)-1}\right)^2\label{5.15}\eeq
where $S_l$ and $T_l$ are (complex) constants of $O(1)$.
It turns out that the quantity $R_l/\sqrt{\wbar_l}$ is never pure imaginary 
 and thus $Re S_l\ne 0$; except for the single real root, $S_l$ is not 
purely real either and hence the correlation functions have some
oscillatory behaviour which is typical of a non-unitary theory.
The formula \rf{5.15} is valid for $\C J\to\infty$; expanding we get
\beq w=\wbar_l\left(1+\sum_{n=1}^\infty\psi^{(l)}_n
\exp\left(-n\Delta^{1-3/2K}\tilde S_l\C J\right)\right)\label{5.16}\eeq
where
\bea \tilde S_l&=& S_l,\quad\mbox{ if $Re S_l>0$,}\nn\\
&=& -S_l,\quad\mbox{ if $Re S_l<0$.}\label{5.16a}\eea
Using \rf{5.10} and \rf{5.16} we see that the conditions \rf{4.65} can be written in the form
\bea 0&=&\frac{1}{s} (z_c+\epsilon_K\Delta^{1/K}\wbar_l)^{m/2}+\sum_{n=1}^\infty
\frac{A^{(n)}_l}{s+n\Delta^{1-3/2K}S_l}\nn\\
&&+2Kg_K^K\left(\frac{1}{s}\left[\mymatrix{M}^{(0)}\BS G\right]_l 
+\sum_{n=1}^\infty   \frac{\left[\mymatrix{M}^{(n)}\BS G\right]_l}{s+n\Delta^{1-3/2K}S_l} \right)\label{5.17}\eea
where $A^{(n)}_l$ and $\mymatrix{M}^{(n)}$ are collections of coefficients and
\beq \mymatrix{M}^{(0)}_{pq}=(z_c+\epsilon_K\Delta^{1/K}\wbar_p)^{K-1-q}.
\label{5.18}\eeq
The apparent singularity in \rf{5.17} at $s=0$ is of course not present 
provided $
\mymatrix{M}^{(0)}$ is invertible.  $
\mymatrix{M}^{(0)}$ is of the Vandermonde form  and therefore
\beq \det \mymatrix{M}^{(0)}=(\epsilon_K\Delta^{1/K})^{(K-1)(K-2)/2}
\prod_{p>q}(\wbar_p-\wbar_q).\label{5.19}\eeq
Since none of the roots are degenerate $\mymatrix{M}^{(0)}$ is invertible.
Therefore the first non-analyticity occurs at $s\sim -\Delta^{1-3/2K}$
and hence the mass gap exponent is
\beq \nu=1-\frac{3}{2K}\label{5.20}\eeq

\section{Discussion}

We have found that $\nu=1-3/2K$ and $\eta =2+2/(2K-3)$ and therefore,
 since $\gamma =-1/K$,
  the Fisher scaling relation \rf{1.F} is indeed satisfied. The results show that the functions which are initially undetermined in the peeling calculation 
do not in fact change the conclusions one would draw simply by ignoring
their contribution in \rf{4.3}.  The value of $\eta$ agrees with that obtained by slicing \cite{klebanovgubser}.

 To compare $\nu$
it is necessary to examine the continuum limit for the perturbation \rf{5.1} away from the multi-critical point.
The multi-critical models have $K$ independent coupling constants so there are
$K$ independent directions along which the multi-critical point can be approached; if chosen appropriately these directions correspond in the continuum
limit to operators of (length) dimensions 
$\{1,2,3,\ldots K\}$ \cite{mooreetal}.
 The procedure for taking the continuum limit
is explained in \cite{book}. 
 By 
definition boundary loops with $l$ legs in the dual graph
(ie which are $l$ links long) have continuum length $L=la$ where $a$ is the 
length of one link. This implies that the generating function variable $x$ conjugate to $l$ must be related to a continuum quantity 
$X$ by $x=x_ce^{-aX}$ where $x_c$ is the radius of convergence of the
disk amplitude \rf{5.6}. We can construct
a non-trivial continuum disk amplitude $A_c(\widetilde\Lambda,X)$ provided that
\beq\Delta\sim \widetilde\Lambda a^K\label{7.A}\eeq
 where $\widetilde\Lambda$ is some 
continuum coupling; we obtain
\beq A(\Delta,x)=A_0(\Delta,x)-a^{K-1/2}A_c(\widetilde\Lambda,X)+\ldots
\label{7.1}\eeq
where $A_0$ is the analytic (non-universal) part of the disk amplitude.
Note that $\widetilde\Lambda$ is not the cosmological constant; in the 
continuum theory the volume $V$ must have dimension $({\rm length})^2$
and therefore the cosmological constant $\Lambda$ must correspond to whichever
direction in the space of lattice couplings leads in the continuum limit to an 
operator of dimension 2. It is straightforward to check that this is 
accomplished by the modified couplings
\bea\bar g^K_p&=&g^K_p-\bar\Delta(-1)^{p-1}\frac{(K-2)!(p-1)!}{(K-p-2)! 2p!},
\quad p\le K-2\nn\\
\bar g^K_p&=&g^K_p,\quad p=K,K-1
.\label{7.B}\eea
The weight for polygons of $2p$ sides is modified by a $p$-dependent
factor and so  the lattice quantity
whose continuum limit is the volume is a complicated object with 
polygons of different number of sides weighted in different ways; the
volume is not the number of polygons.

Using the scaling \rf{7.A} for $\Delta$ in the large $t$ behaviour of the two-point function we find that 
\beq \exp(-\Delta^\nu t)=\exp(-\widetilde\Lambda^\nu(ta^{K\nu}))=
\exp(-\widetilde\Lambda^\nu T)\label{7.2}\eeq
and hence we will get a consistent non-trivial scaling limit 
provided  the continuum string time scales as $T\sim t a^{K\nu}$.
This also agrees with  \cite{klebanovgubser} (where what we call the 
string time is called the geodesic distance).
 
The structure of these multi-critical surfaces seems slightly bizarre.
Recall that  the coupling $\Delta$ is conjugate to the number of polygons
which therefore behaves roughly like $\Delta^{-1}$ for typical surfaces
in the ensemble.  On the other hand from \rf{7.2} we have that the typical
$t$ must be roughly $\Delta^{-\nu}$ and therefore
\beq \expect{ \#\hbox{polygons}}\sim \expect{t}^\frac{1}{\nu}.\label{7.3}\eeq
When $K$ gets large $\nu\to 1$ and so these surfaces have an almost linear
structure when viewed in terms of polygons.  However, as we discussed in section 1, the polygons can be very large so that, for example, the number of links
traversed in a typical cycle can also be very large and it does not follow that the surfaces in the continuum limit are at all one-dimensional.  It does however cast doubt on the idea that it makes sense to identify the string time $T$
with a continuum geodesic distance.

Another manifestation of this can be seen by  considering a perturbation in terms of the cosmological constant $\Lambda$. Then
 we  expect that the two
point function behaves as
\beq \exp(-\Lambda^{\frac{K\nu}{2}} T)\label{7.4}\eeq
(it is a straightforward calculation along the lines of section 5 but using 
the couplings \rf{7.B} to check
this).
The average volume of the system will be $\expect{V}\sim \Lambda^{-1}$;
on the other hand from \rf{7.4} the string time extent  must be of order
$T_s\sim\Lambda^{-\frac{K\nu}{2}}$
\renewcommand{\thefootnote}{\fnsymbol{footnote}}
\footnote[2]{One might worry that the negative weights appearing in the
multi-critical models cause some cancellation which alters this conclusion. We have checked explicitly that for $K=4$ at least this does not happen.}
 and hence
$\expect{V}\sim T_s^{\frac{2}{K\nu}}$. Interpreting $T$ as a geodesic 
distance leads to the conclusion that the Hausdorff  dimension is $d_H=\frac{2}{K\nu}$ which vanishes as $K\to\infty$ and is less than two for all  $K>2$.
However we can look at this another way; in \cite{klebanovgubser} the probability distribution for the length $L$ of exit loops at time $T$ given a point-like entrance
loop at $T=0$ was calculated. At large but finite $K$ the distribution shows that there is typically one macroscopic exit loop of length $L\sim T^{\frac{1}{K\nu}}$ but that as $K\to\infty$ the distribution function becomes a delta-function. Using this result we can relate the volume to the typical boundary length
as $K\to\infty$ and find
\beq V\sim L^2\eeq
which is typical of a smooth flat surface of Hausdorff dimension 2. This is the behaviour we expect because $K\to\infty$ corresponds to central charge $c\to-\infty$ where semi-classical properties are recovered; the volume and loop
length have highly anomalous behaviour relative to the string time but exactly what we expect relative to each other.  Note that our discussion
above implies that the macroscopically large boundary loops are often made from a finite number of polygons with diverging number of sides.

The multi-critical models are non-unitary and have negative central charge,
and so it is plausible that in the continuum theory the Hausdorff
 dimension is given by
\rf{1.7} which works well for the $c=-2$ model.
As $c\to-\infty$ \rf{1.7} gives $d_H=2$ and so it is clear that the 
geodesic distance implicit in the derivation of this formula (which is done directly in the continuum using scaling arguments and Liouville theory)
cannot be equivalent to the continuum limit of the string time, $T$.
Since we do not know what the relationship between these distance measures
really is, it is not clear what calculation in the discretized formulation
would be needed to check \rf{1.7} for general $K$; in any case it must be 
highly non-trivial because \rf{1.7} gives irrational values for $d_H$ when 
$K>2$ whereas discretized calculations are almost sure to give rational
values for exponents.

\vskip1cm
We acknowledge useful conversations with Thordur Jonsson. This work 
was supported in part by PPARC Grant GR/L56565.

\end{document}